\documentclass[aps,prd,twocolumn,groupedaddress,nofootinbib,showpacs]{revtex4}%
\usepackage{graphicx}

\usepackage{amsmath}
\usepackage{amssymb}
\usepackage{multirow}
\usepackage{bm}
\usepackage{color}

\usepackage[hypertex]{hyperref}

\begin{document}
\title{Impact of $J/\psi$ pair production at the LHC and predictions in nonrelativistic QCD\\[7mm]}

\author{Li-Ping Sun$^a$}
\email{sunliping@pku.edu.cn}
\author{Hao Han$^a$}
\email{hao.han@pku.edu.cn}
\author{Kuang-Ta Chao$^{a,b,c}$}
\email{ktchao@pku.edu.cn}

\affiliation{ {\footnotesize (a)~School of Physics and State Key
Laboratory of Nuclear Physics and Technology, Peking University,
Beijing 100871, China}\\{\footnotesize (b)Collaborative Innovation Center of Quantum Matter, Beijing, China}\\
{\footnotesize (c)~Center for High Energy physics, Peking
University, Beijing 100871, China}}

\begin{abstract}
For $J/\psi$ pair production at hadron colliders, we present the
full next-to-leading order (NLO) calculations with the color-singlet
channel in nonrelativistic QCD. We find that the NLO
result can reasonably well describe the LHCb measured cross section, but
exhibits very different behaviors from the CMS data in the
transverse momentum distribution and mass distribution of $J/\psi$
pair. Moreover, by adding contributions of gluon fragmentation and quark
fragmentation, which occur at even higher order in $\alpha_s$, it
is still unable to reduce the big differences. In particular, the observed
flat distribution in the large invariant mass region is hard to
explain. New processes or mechanisms are needed to understand
the CMS data for $J/\psi$ pair production.
\end{abstract}
\maketitle

\subsection{INTRODUCTION}

Nonrelativistic QCD (NRQCD)\cite{nrqcd} is
widely used in the study of heavy quarkonium physics. In NRQCD a
quarkonium production process can be factorized as short-distance parton scattering amplitudes  multiplied by long-distance
matrix elements (LDMEs). This factorization has been applied
in single quarkonium production and tested by various
experiments\cite{Inc1,Inc2,Inc3,Inc4,Inc5,Inc6}.

Besides the single quarkonium production, the multi-quarkonuim
production provides another ideal laboratory to understand the
quarkonium production mechanism that NRQCD assumes. At the LHC, the
LHCb Collaboration in 2011 measured the $J/\psi$ pair production for
the first time at the center-of-mass energy
$\sqrt{s}=7~\mathrm{TeV}$ with an integrated luminosity of
$35.2~\mathrm{pb}^{-1}$\cite{LHCb}. In 2013, the CMS Collaboration
further released the data of $J/\psi$ pair production\cite{CMS} with
a much larger transverse moment range, providing a good platform for
testing the validity of NRQCD in quarkonium pair production.

In Refs.\cite{LO1,LO2,LO3}, the leading order (LO) calculation of
$J/\psi$ pair production in the color singlet model (CSM) is
performed. The relativistic correction to the $J/\psi$ pair
production is carried out in Ref.\cite{RC}, where the relativistic
correction makes significant improvement for diluting the
discrepancy between the shapes of color-singlet (CS) and color-octet
(CO) differential cross sections at LO. Furthermore, the partial
next-to-leading order ($\mathrm{NLO}^{\star}$) correction for
$J/\psi$ pair production is evaluated by Lansberg and Shao
\cite{NLOstar}. They argue that the $\mathrm{NLO}^{\star}$ yield can
approach the full NLO result at large $p_T$,  the transverse
momentum of one of the two $J/\psi$'s, and thus the
$\mathrm{NLO}^{\star}$ results give a more precise theoretical
prediction than the LO results in this region. 
All the above works are performed in the single
parton scattering (SPS) mechanism, while the contribution of double
parton scattering (DPS) is assessed in Refs.\cite{DPS1,DPS2,DPS3},
and is expected to be important. As predictions for DPS are very
model-dependent \cite{DPS1,DPS2,DPS3}, it is needed to have an
accurate calculation for SPS contribution before one can extract the
DPS contribution.

In order to further understand the multi-quarkonium production
mechanism, it is necessary to evaluate the $J/\psi$ pair production
at NLO, which is the main work in this paper. Compared to the LO
calculation, the NLO calculation is expected to not only reduce the theoretical
uncertainties, but also open new kinematic enhanced topologies,
which may dominate at large $p_T$. More precisely, we may find
that at NLO the differential cross section $d\sigma/dp_T^2$ at large $p_T$
behaves as $p_T^{-6}$  due to double parton fragmentation contributions \cite{DPF}, while it only behaves as $p_T^{-8}$ at LO. Moreover, we also
include the dominant $p_T^{-4}$ contribution via single parton
fragmentation, which contributes at even higher order in $\alpha_s$ and also involves color-octet channels.
Thus we will obtain the most precise predictions for $J/\psi$ pair production with the color-singlet channel as well as some color-octet effects in the fragmentation contributions.

\subsection{FORMULISM}

In NRQCD, the cross section of $J/\psi$ pair
production at the LHC can be expressed as \cite{nrqcd}
\begin{eqnarray}
d\sigma_{p+p \to J/\psi+J/\psi}=\sum_{i,j,n_1,n_2}{\int}dx_1dx_2{f_{i/p}(x_1)}{f_{j/p}(x_2)} \nonumber\\
\times~{d\hat{\sigma}^{n_1,n_2}_{i,j}}\langle\mathcal{O}_{n_1}\rangle^{J/\psi}
\langle\mathcal{O}_{n_2}\rangle^{J/\psi}. \label{eq:factorization}
\end{eqnarray}
where $f_{i/p}(x_{1,2})$ are the parton distribution functions
(PDFs), $x_1$ and $x_2$ represent the momentum fraction of initial
state partons from the protons,
$\langle\mathcal{O}_{n}\rangle^{J/\psi}$ are LDMEs of $J/\psi$ with
$n = {}^{2S+1}L_J^{[c]}$ are the standard spectroscopic notation for
the quantum numbers of the produced intermediate heavy quark pairs,
and $d\hat{\sigma}$ are partonic short distance coefficients.
For the $J/\psi$ pair production we usually set
$n_1=n_2={}^{3}S_1^{[1]}$ in Eq.~\eqref{eq:factorization} but other
intermediate states may also be specified.

\begin{figure}[!hbtp]
\centering
\includegraphics[scale=0.45]{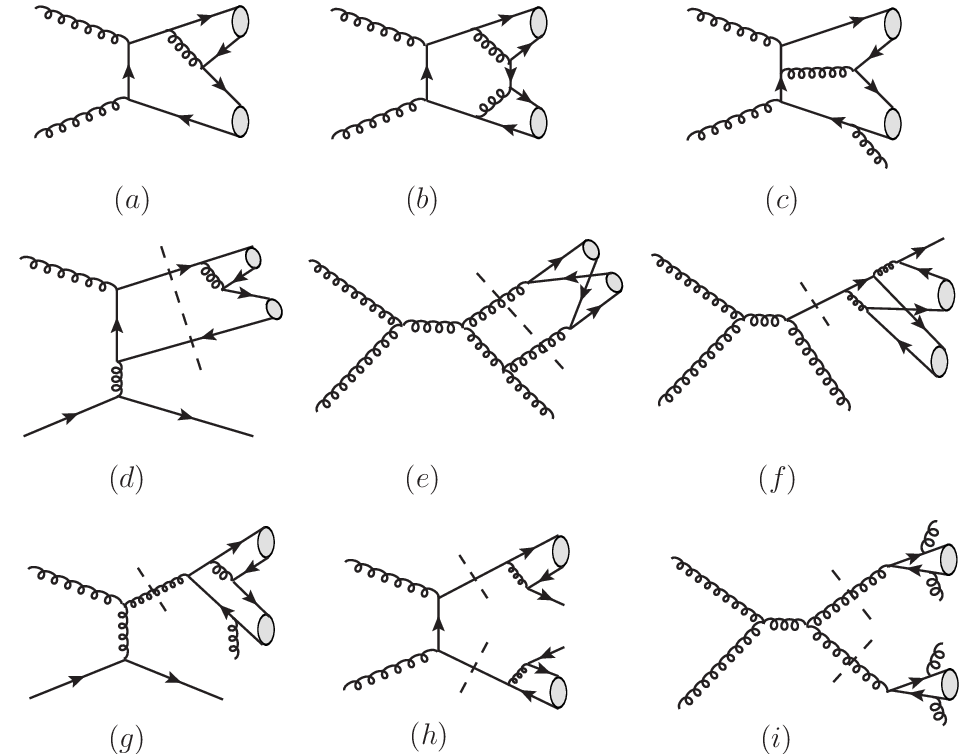}%
\caption{\small Typical Feynman diagrams for $J/\psi$ pair
production in color-singlet channel, including LO, NLO, as well as
single quark or gluon fragmentation diagrams beyond NLO.}
\label{feynmandiag}
\end{figure}

In the LO calculation, there are two subprocesses:
$g+g{\rightarrow}J/\psi+J/\psi$ and
$q+\bar{q}{\rightarrow}J/\psi+J/\psi$, only the former is taken into
account since the contribution of the other process is highly
suppressed by the quark PDFs. While in the NLO case, besides the
gluon fusion process, the quark gluon process $q+g\rightarrow
2J/\psi+q$ should also be considered. Typical Feynman diagrams at LO
and NLO are shown in Fig.\ref{feynmandiag} $(a)-(c)$.

The $c\bar{c}$ pair hadronization process can be computed by using
the covariant projection operator method, for $J/\psi(^{3}S_{1})$,
we employ the following commonly used projection operators for spin
and color:
\begin{eqnarray}
\Pi_{1} = \frac{1}{\sqrt{8m_c^3}}\left(\frac{\not\!P}{2}-m_c\right)
\not\!\epsilon_{J/\psi}\left(\frac{\not\!P}{2}+m_c\right).\label{eq:6}
\end{eqnarray}
and
\begin{eqnarray}
\mathcal{C}_1=\frac{1}{\sqrt{N_c}}. \label{eq:5}
\end{eqnarray}
where $\epsilon^{\mu}_{J/\psi}$ is the $J/\psi$ polarization vector
with $P\cdot \varepsilon=0$, $P$ is the momentum of $J/\psi$.


The NLO contributions can be divided into two parts: the virtual
correction and the real correction. The virtual correction which
arises from loop diagrams includes gluon fusion process only, the
same as the LO case, while for the real correction, besides the
gluon fusion process, the process $q+g\rightarrow 2J/\psi+q$ should
also be taken into account.

In the virtual correction, the ultraviolet(UV) and infrared(IR)
divergences usually exist. We use the dimensional regularization
scheme to regularize the UV and IR divergences. The Coulomb
divergence caused by the virtual gluon line connecting the quark pair
in a $J/\psi$, is regularized by the relative velocity $v$. The UV
divergences can be renormalized by counter terms. The
renormalization constants include $Z_{2}$, $Z_{3}$, $Z_{m}$, and
$Z_{g}$, corresponding to quark field, gluon field, quark mass, and
strong coupling constant $\alpha_{s}$, respectively. Here, in our
calculation the $Z_{g}$ is defined in the
modified-minimal-subtraction ($\mathrm{\overline{MS}}$) scheme,
while for the other three the on-shell ($\mathrm{OS}$) scheme is
adopted, which reads
\begin{eqnarray}
\delta Z_{m}^{OS} &=&-3C_{F}\displaystyle\frac{\alpha _{s}}{4\pi }\left[ %
\displaystyle\frac{1}{\epsilon _{UV}}-\gamma _{E}+\ln \displaystyle\frac{%
4\pi \mu _{r}^{2}}{m_{c}^{2}}+\frac{4}{3}\right] ,  \nonumber \\
\delta Z_{2}^{OS} &=&-C_{F}\displaystyle\frac{\alpha _{s}}{4\pi }\left[ %
\displaystyle\frac{1}{\epsilon _{UV}}+\displaystyle\frac{2}{\epsilon _{IR}}%
-3\gamma _{E}+3\ln \displaystyle\frac{4\pi \mu
_{r}^{2}}{m_{c}^{2}}+4\right]
,  \nonumber \\
\delta Z_{2l}^{OS} &=&-C_{F}\displaystyle\frac{\alpha _{s}}{4\pi }\left[ %
\displaystyle\frac{1}{\epsilon _{UV}}-\displaystyle\frac{1}{\epsilon _{IR}}%
\right] ,  \nonumber \\
\delta Z_{3}^{OS} &=&\displaystyle\frac{\alpha _{s}}{4\pi }\left[
(\beta_{0}^{'}-2C_{A})\left( \displaystyle\frac{1}{\epsilon _{UV}}-\displaystyle\frac{1%
}{\epsilon _{IR}}\right)\right.\nonumber\\
&-&\left. \frac{4}{3}T_f(n_f-n_{lf})\left(\frac{1}{\epsilon _{UV}}-\gamma_E+\ln \displaystyle\frac{%
4\pi \mu _{r}^{2}}{m_{c}^{2}}\right)\right] , \nonumber\\
\delta Z_{g}^{\overline{\mathrm{MS}}} &=&-\displaystyle\frac{\beta _{0}}{2}%
\displaystyle\frac{\alpha _{s}}{4\pi }\left[
\displaystyle\frac{1}{\epsilon _{UV}}-\gamma _{E}+\ln (4\pi )\right]
.\label{eq:9}
\end{eqnarray}
where $\beta _{0}=\frac{11}{3}C_{A}-\frac{4}{3}T_{F}n_{f}$ is the
one-loop coefficient of the QCD beta function; $n_{f}=4$ is the
number of active quarks in our calculation; $\beta
_{0}^{'}=\frac{11}{3}C_{A}-\frac{4}{3}T_{F}n_{lf}$ with $n_{lf}=3$
the number of light quarks. $C_{A}=3$ and $T_{F}=1/2$ attributed to
the SU(3) group; $\mu_r$ is the renormalization scale.

As mentioned above, there are two processes involved in the real
corrections: $g+g\rightarrow J/\psi+J/\psi+g$ and $q+g\rightarrow
J/\psi+J/\psi+q$. It is known that IR divergence exists in these
processes because of the phase space integration, which can be
canceled by the IR sigularities left in the virtual correction.
According to the different regions of the phase space, the IR
divergence can be categorized as soft or collinear. In this paper,
we use the two-cutoff phase space slicing method\cite{twocut} to
isolate the two types of IR sigularities, then the cross section of
real correction can be expressed as:
\begin{eqnarray}
\sigma_{Real}=\sigma_{Real}^{Soft}+\sigma_{Real}^{HC}+\sigma_{Real}^{\overline{HC}}.
\label{eq:11}
\end{eqnarray}
where $HC$ and $\overline{HC}$ represent hard collinear and hard
non-collinear contributions, respectively.

The soft sigularities only originate from real gluon emission, that
is, the $g(p_1)+g(p_2)\rightarrow J/\psi(p_3)+J/\psi(p_4)+g(p_5)$
process. $p_5$ is the momentum of the emitted gluon, and in the
$p_1+p_2$ rest frame, $p_1+p_2=\sqrt{s_{12}}(1,0,0,0)$. Applying the
two cutoff technique, the soft region is defined in the $p_1+p_2$
rest frame by $0\leq E_5\leq \delta_s \sqrt{s_{12}}/2$, where $
\delta_s$ is a small cut.

In the soft region, the three-body phase space can be simplified as:
\begin{eqnarray}
d\mathrm{PS}_3|_{Soft} &=& d\mathrm{PS}_2
\frac{d^{d-1}p_5}{2p^0_5(2\pi)^{d-1}}|_{Soft}\nonumber\\
&=&d\mathrm{PS}_2 \left[ \left( \frac{4\pi}{s_{12}} \right)^\epsilon
\frac{\Gamma(1-\epsilon)}{\Gamma(1-2\epsilon)} \frac{1}{2(2\pi)^2}
\right] \mathrm{dS}.\nonumber\\
\label{eq:13}
\end{eqnarray}
with
\begin{eqnarray}
\mathrm{dS} &=& \frac{1}{\pi}
     \left( \frac{4}{s_{12}} \right)^{-\epsilon}\int_0^{\delta_s\sqrt{s_{12}}/2}
     dE_5 E_5^{1-2\epsilon}\nonumber\\
     &&\times \sin^{1-2\epsilon}\!\theta_1\,d\theta_1\sin^{-2\epsilon}\!\theta_2\,d\theta_2 \, .
\label{eq:14}
\end{eqnarray}

Meanwhile, the relative matrix elements in the soft region can
be factorized as
\begin{eqnarray}
M^a_3|_{Soft} \simeq
g\mu_r^{\epsilon}\varepsilon^{\mu}(p_5)\mathbf{J}_{\mu}^a(p_5)
\mathbf{M}_2, \label{eq:15}
\end{eqnarray}
where $a$ is the color index the emitted gluon carries, and
$\varepsilon^{\mu}(p_5)$ is the gluon's polarization vector.
$\mathbf{M}_2$ is the color connected LO Born matrix element,
$\mathbf{J}_{\mu}^a(p_5)$ is the non-abelian eikonal current, which
contains the color structure of the emitted gluon and the soft
divergence information. The concrete form of
$\mathbf{J}_{\mu}^a(p_5)$ is given by:
\begin{eqnarray}
\mathbf{J}_{\mu}^a(p_5) = \sum_{f} \mathbf{T}^a_f \frac{p_f}{p_f
\cdot p_5},\label{eq:16}
\end{eqnarray}
where the sum goes over each external line that can emit a soft
gluon, the color structure associated with each soft gluon emission
from parton $f$ is denoted by $\mathbf{T}_f$. Then the squared
matrix element reads:
\begin{eqnarray}
|M_3|^2|_{soft} \simeq -g^2\mu_r^{2\epsilon} \sum_{f,{f^\prime}}
\frac{p_f \cdot p_{f^\prime}}{p_f \cdot p_5 \; p_{f^\prime} \cdot
p_5} M^0_{f{f^\prime}},\label{eq:17}
\end{eqnarray}
with
\begin{eqnarray}
M^0_{f{f^\prime}} &=& ( \mathbf{T}^a_f \mathbf{M}_2 ) (
\mathbf{T}^a_{f^\prime}
    \mathbf{M}_2 )\nonumber\\
  &=& \left[ M_{c_1 \ldots b_f \ldots b_{f^\prime} \ldots c_4} \right]^*
    T_{b_f d_f}^a T_{b_{f^\prime} d_{f^\prime}}^a
    M_{c_1 \ldots d_f \ldots d_{f^\prime} \ldots c_4}.\nonumber\\
    \label{eq:18}
\end{eqnarray}

Combining the phase space and squared matrix element given above,
one can finally get the cross section of real correction in the soft
region:
\begin{eqnarray}
d\sigma_{Real}^{Soft} &=&
         \left[ \frac{\alpha_s}{2\pi} \frac{\Gamma(1-\epsilon)}{\Gamma(1-2\epsilon)}
         \left( \frac{4\pi\mu_r^2}{s_{12}} \right)^\epsilon \right]
         \sum_{f,{f^\prime}} d\sigma_{f{f^\prime}}^{Born}\nonumber\\
         &&\times\int
         \frac{-p_f \cdot p_{f^\prime}}{p_f \cdot p_5 \; p_{f^\prime}
         \cdot p_5} \mathrm{dS} \, ,
\label{eq:19}
\end{eqnarray}
with
\begin{eqnarray}
d\sigma^{Born}_{f{f^\prime}} \propto \overline{\sum}
M^{0}_{f{f^\prime}} d\mathrm{PS}_2 \, . \label{eq:20}
\end{eqnarray}

We can see in Eq.~\eqref{eq:19} that in the soft region, the
divergence is singled out. All the concrete expressions of the
integration $ \int \frac{-p_f \cdot p_{f^\prime}}{p_f \cdot p_5 \;
p_{f^\prime} \cdot p_5} \mathrm{dS}$ are listed in the Appendix of
Ref.\cite{twocut}.

The hard collinear divergence only occurs at massless case, so it is
also called ``mass singularity". According to the two cutoff method,
a small cut $\delta_c$ is brought in, and the hard collinear region
of the phase space is that where any invariants ($s_{ij}$ or
$t_{ij}$) gets smaller than $\delta_c s_{12}$. The hard-collinear
divergence can be divided into initial state collinear and final
state collinear, depending on the singularities from initial or
final state. For our process, there is only initial state collinear
because the $J/\psi$ pair in the final state are massive. The
processes include: $g(p_1)+g(p_2)\rightarrow
J/\psi(p_3)+J/\psi(p_4)+g(p_5)$ and $g(p_1)+q(p_2)\rightarrow
J/\psi(p_3)+J/\psi(p_4)+q(p_5)$. Hereafter, we only consider the
case that the emitting and splitting occur at parton $g(p_2)$ and
$q(p_2)$, that is, $0\leq t_{25}=(p_2-p_5)^2\leq \delta_c s_{12}$,
the other cases are tackled the same way.

In the hard-collinear region, the three-body phase space can be
written as:
\begin{eqnarray}
&&d\mathrm{PS}_3|_{HC}=\bigg[\frac{d^{d-1}p_3}{2p_3^0(2\pi)^{d-1}}\frac{d^{d-1}p_4}
{2p_4^0(2\pi)^{d-1}}\nonumber\\
&&~\times(2\pi)^d d^d(p_1+zp_2-p_3-p_4)\bigg]
\frac{d^{d-1}p_5}{2p_5^0(2\pi)^{d-1}}.\label{eq:21}
\end{eqnarray}
where $z$ is the momentum fraction for the splitting $2\rightarrow
2{'}+5$, by applying the collinear approximation, the three-body
matrix elements can be expressed as follows:
\begin{eqnarray}
&&\overline{\sum} \vert M_3(1+2\rightarrow 3+4+5)\vert^2\nonumber\\
& \simeq &\overline{\sum} \vert M_2( 1+2'\rightarrow 3+4)\vert^2
P_{2'2}(z,\epsilon)g^2 \mu_r^{2\epsilon} \frac{-2}{z
t_{25}}.\nonumber\\
\label{eq:22}
\end{eqnarray}

Combining the phase space and the matrix elements, we can obtain the
cross section in the hard collinear region:
\begin{eqnarray}
&&d\sigma_{Real}^{HC}(p+p\rightarrow 2J/\psi+X)\nonumber\\
& = &\sum_{i=g,q} f_{g/p}(x_1)f_{i/p}(x_2/z)\left[
\frac{\alpha_s}{2\pi} \frac{\Gamma(1-\epsilon)}{\Gamma(1-2\epsilon)}
\left(\frac{4\pi\mu_r^2}{s_{12}}\right)^{\epsilon}\right] \nonumber \\
&&~\times d\hat\sigma_0^{Born}\left(-\frac{1}{\epsilon}\right)
\delta_c^{-\epsilon}P_{gi}(z,\epsilon)\frac{dz}{z} \left[ \frac
{(1-z)}{z} \right]^{-\epsilon}dx_1dx_2.\nonumber\\
\label{eq:23}
\end{eqnarray}

The collinear singularity emerging in Eq.~\eqref{eq:23} should be
factorized into the parton distribution functions. To do this, a
scale dependent parton distribution function is introduced using the
$\overline{\mathrm{MS}}$ convention:
\begin{eqnarray}
f_{b/B}(x,\mu_f)&=&f_{b/B}(x)-\frac{1}{\epsilon} \left[
\frac{\alpha_s}{2\pi} \frac{\Gamma(1-\epsilon)}{\Gamma(1-2\epsilon)}
\left(\frac{4\pi
\mu_r^2}{\mu_f^2}\right)^{\epsilon}\right]\nonumber\\
&&\times\int_z^1 \frac{dz}{z} P_{bb'}(z)f_{b'/B}(x/z). \label{eq:24}
\end{eqnarray}

After renormalization of the parton distribution function, we
can eventually obtain the cross section for the initial state
collinear contribution:
\begin{eqnarray}
&&d\sigma_{Real}^{HC}(p+p\rightarrow 2J/\psi+X)\nonumber\\
&=&d\hat{\sigma}_{0}^{Born}\left[\frac{\alpha_s}{2\pi}\frac{\Gamma(1-\epsilon)}{\Gamma(1-2\epsilon)}(\frac{4\pi\mu_r^2}{s_{12}})^\epsilon\right]\nonumber\\
&&\times\bigg\{f_{g/p}(z_1,\mu_{f})\tilde{f}_{g/p}(z_2,\mu_{f})+\bigg[\frac{A_1^{SC}(g\rightarrow
g+g)}{\epsilon}\nonumber\\
&&+A_0^{SC}(g\rightarrow g+g)\bigg]
f_{g/p}(z_1,\mu_{f})f_{g/p}(z_2,\mu_{f})\bigg\}dz_1dz_2.\nonumber\\
\label{eq:25}
\end{eqnarray}

Note that in this expression, the collinear singularity is absorbed
into the redefinition of the parton distribution function. The left
soft collinear factors $A_i^{SC}$ result from the difference of the
upper bound of the $z$ integration in Eq.(\ref{eq:23}) and
Eq.(\ref{eq:24}). These factors are given by:
\begin{eqnarray}
A_1^{SC}(g\rightarrow g+g)&=&2 N \ln \delta_s+(11N-2n_f)/6,\nonumber\\
A_0^{SC}(g\rightarrow g+g)&=&\left[2 N \ln
\delta_s+(11N-2n_f)/6\right]\ln(\frac{s_{12}}{\mu_f^2}).\nonumber\\
\label{eq:26}
\end{eqnarray}
There is no $A_i^{SC}(q\rightarrow q+g)$ term existing because the
$q\rightarrow q+g$ splitting process demonstrates no soft
singularities. The $\tilde{f}$ functions read:
\begin{eqnarray}
\tilde{f}_{g/p}(z,\mu_{f})=\sum_{i}\int_{z}^{1-\delta_s\delta_{gi}}\frac{dy}{y}
~~f_{i/p}(\frac{z}{y},\mu_{f})~~\tilde{P}_{gi}(y),\label{eq:27}
\end{eqnarray}
with
\begin{eqnarray}
\tilde{P}_{ij}(y)=P_{ij}(y)~\ln\left(\delta_c\frac{1-y}{y}\frac{s12}{\mu_{f}^{2}}\right)-P^{'}_{ij}(y).
\label{eq:28}
\end{eqnarray}
where the index $i$ in the sum represents a gluon or a quark, and
the d-dimension unregulated splitting functions $P_{ij}(y)$ and
$P^{'}_{ij}(y)$ are given by:
\begin{eqnarray}
P_{qq}(y)&=&C_F\frac{1+y^2}{1-y},\nonumber\\
P^{'}_{qq}(y)&=&-C_F(1-y),\nonumber\\
P_{gq}(y)&=&C_F\frac{1+(1-y)^2}{y},\nonumber\\
P^{'}_{gq}(y)&=&-C_Fy,\nonumber\\
P_{gg}(y)&=&2\mathrm{N}\left[\frac{y}{1-y}+\frac{1-y}{y}+y(1-y)\right],\nonumber\\
P^{'}_{gg}(y)&=0&,\nonumber\\
P_{qg}(y)&=&\frac{1}{2}[y^2+(1-y)^2],\nonumber\\
P^{'}_{qg}(y)&=&-y(1-y).
\label{eq:29}
\end{eqnarray}

Now, the cross sections for the $J/\psi$ pair production at NLO
can be expressed as:
\begin{eqnarray}
\sigma_{NLO}=\sigma_{Born}+\sigma_{Virtual}+\sigma_{Real}.
\label{eq:30}
\end{eqnarray}

The soft divergences and collinear divergences from real corrections
will cancel divergences from virtual corrections, and thus the final
NLO contributions are IR safe.

Because there are two $J/\psi$ states in the final state, the LO
contributions behave as $p_T^{-8}$ when $p_T$ is large. However, at
NLO level, there are double quark and double gluon fragmentation
contributions [Fig.~\ref{feynmandiag} (d) and (e)], which give
$p_T^{-6}$ behavior \cite{DPF}. We thus expect that the NLO
contribution will dominate at large $p_T$, especially for the CMS
data, where a relatively large lower $p_T$ cutoff is
taken\cite{CMS}. Since in the double parton fragmentation diagrams
the two $J/\psi$'s come from the same fragmenting partons, the
invariant mass of the pair (denoted as $M_{J/\psi J/\psi}$) should
be small. This implies that the NLO correction will be significant
only in the small $M_{J/\psi J/\psi}$ region, and it will be mild
when $M_{J/\psi J/\psi}$ is large.
All these expectations will be confirmed by our numerical results
shown below.

When $p_T$ is large enough, the single parton fragmentation
contributions, which behave as $p_T^{-4}$, will eventually dominate,
although they are suppressed by powers of $\alpha_s$. For double
$J/\psi$ production, the quark and gluon fragmentation processes can
be expressed as
\begin{eqnarray}
d \sigma_{A+B \to 2J/\psi + X}=\sum_{i,j,n_1,n_2}
d\hat\sigma_{A+B\to
i+j+X}\nonumber\\
\otimes D_{i\to Q \bar Q(n_1)}\otimes D_{j\to Q \bar Q(n_2)}
\langle\mathcal{O}_{n_1}\rangle \langle\mathcal{O}_{n_2}\rangle,
\label{eq:Frag}
\end{eqnarray}
where $D_{i,j\to Q \bar Q(n)}$ are the single-parton fragmentation
functions (FFs) for a NRQCD state $n$. Typical Feynman diagrams for
these kinds of fragmentation contributions are shown in
Fig.\ref{feynmandiag} (h) and (i). These FFs are factorization scale
dependent, and satisfy the DGLAP evolution equation
\cite{DGLAP1,DGLAP2,DGLAP3,DGLAP4,DGLAP5}
\begin{equation}
\frac{d}{d \log \mu_f^2}
\begin{pmatrix}
D_c \\ D_g
\end{pmatrix}
= \frac{\alpha_s(\mu_f)}{2 \pi}
\begin{pmatrix}
P_{cc} & P_{gc} \\
P_{cg} & P_{gg}
\end{pmatrix}
\otimes
\begin{pmatrix}
D_c \\ D_g
\end{pmatrix},
\label{eq:DGLAP}
\end{equation}
where $D_g$ and $D_c$ denote the FFs from gluon and charm quark,
respectively, and $P_{ij}$'s are the splitting functions. Based on
this evolution equation, we only need inputs of FFs at an initial
scale, which can be found in Ref.\cite{FFs}. Note that fragmentation
functions in color-octet channels will also be considered in
Eq.~\eqref{eq:Frag}.

In addition, there are also $p_T^{-4}$ contributions coming from
Feynman diagrams like Fig.\ref{feynmandiag} (f) and (g), where one
parton fragments to a $J/\psi$ pair. We will argue later that these
contributions should not be important.

\subsection{Numerical Inputs}

Because of the complexity of the $J/\psi$
pair production, in our calculation, the package FEYNARTS
\cite{feynarts} is used to generate the Feynman diagrams and
amplitudes. The phase space integration is evaluated by employing
the package Vegas.

In numerical calculation, the CTEQ6L1 and CTEQ6M parton distribution
functions \cite{cteq1,cteq2} are used. The renormalization scale
$\mu_r$ and factorization scale $\mu_f$ are chosen as
$\mu_r=\mu_f=m_T$, with $m_T=\sqrt{p_T^2+16m_c^2}$ and charm quark
mass $m_c=M_{J/\psi}/2=1.55~\mathrm{GeV}$. In the two cutoff method,
there are soft and collinear cutoffs, $\delta_s$ and $\delta_c$,
which we set to be $\delta_s=10^{-2}$ and $\delta_c=10^{-4}$.
Theoretical uncertainties are estimated by varying $\mu_r=\mu_f$
from $m_T/2$ to $2m_T$.

The CS LDME $\langle\mathcal{O}(^3\!S_1^{[1]})\rangle^{J/\psi}=1.16
\rm{GeV}^3$ is estimated by using the $\mathrm{B-T}$ potential
model\cite{BT}. While CO LDMEs for ${}^1S_0^{[8]}$, ${}^3S_1^{[8]}$
and ${}^3P_0^{[8]}$ channels, which are needed in fragmentation
processes, are taken from three different extractions
\cite{chao:2012,wang:2013,kniehl:2012}. Meanwhile, the
${}^1S_0^{[8]}$-dominant CO matrix elements extracted
from\cite{chao:2011} are also taken into account.

\begin{figure}[!hbtp]
\centering
\includegraphics[scale=0.7]{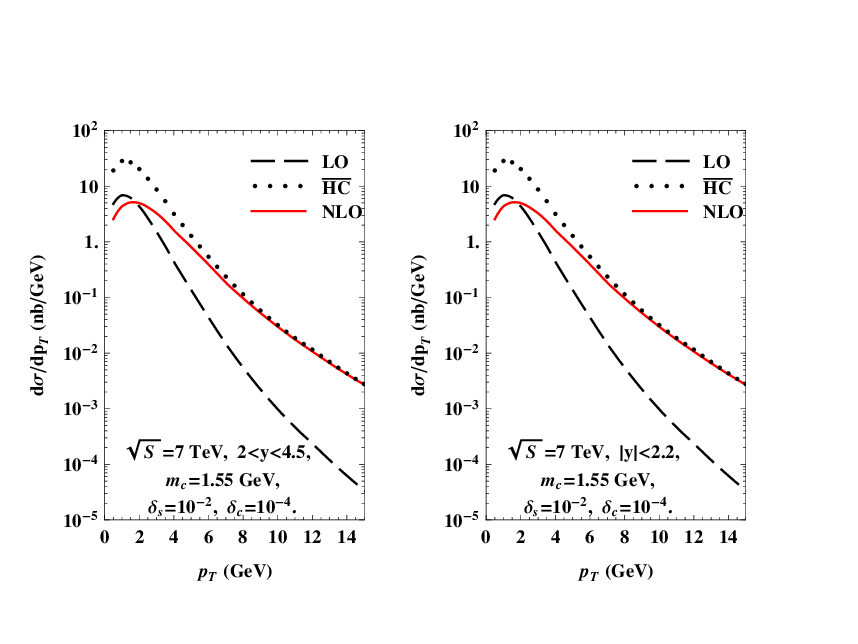}%
\caption{\small (color online). Comparison between LO,
$\overline{HC}$, and full NLO results of the cross section $p_T$ distribution in
$J/\psi$ pair production.} \label{NLONLOstar}
\end{figure}

\subsection{RESULTS}

To see the importance of NLO calculation, we show
the cross section $p_T$ distribution of one of the two $J/\psi$'s in
Fig.~\ref{NLONLOstar} for both forward region and central region in
rapidity. In the low $p_T$ region, although NLO results are close to
LO results, their behaviors are different. Especially, the
NLO result peaks at a larger $p_T$ than that of LO result. When
$p_T\gtrsim 5\mathrm{GeV}$, NLO results become much larger than the
LO one. As emphasized above, the large NLO corrections are due to
the $p_T^{-6}$ contributions from double parton fragmentation. To
demonstrate this point, we show also the hard noncollinear
contributions of real correction $\sigma_{Real}^{\overline{HC}}$,
which contain all the $p_T^{-6}$ contributions, in
Fig.~\ref{NLONLOstar}. As expected, the hard noncollinear
contributions approach the full NLO result as $p_T$ becomes larger. As
for the $\mathrm{NLO}^{\star}$ result in Ref.~\cite{NLOstar}, which
introduces cutoffs to regularize soft and collinear divergences in
the real corrections, it should be similar to our hard
noncollinear contributions. So the $\mathrm{NLO}^{\star}$ result can
give a good approximation to the full NLO result for double $J/\psi$
production in the high $p_T$ region. But the problem of infrared divergence and cutoff dependence at NLO* is removed in our full NLO calculation.

\begin{figure}[!hbtp]
\centering
\includegraphics[scale=0.8]{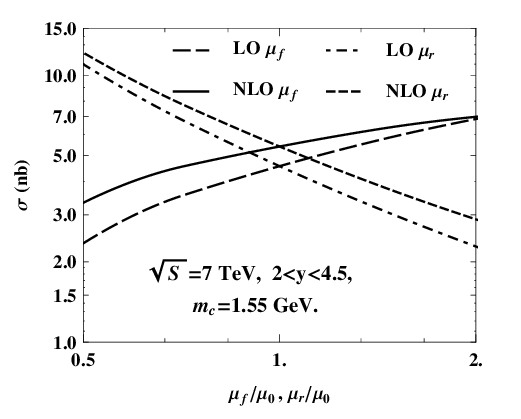}%
\caption{\small  (color online).  Scales dependence of total cross
sections for LO and NLO at LHCb, where $\mu_0=m_T$. }
\label{scaledependence}
\end{figure}

At the LHCb window with $\sqrt{\mathrm{S}}=7~\mathrm{TeV}$,
$2<y(J/\psi)<4.5$, and $0<p_T<10~\mathrm{GeV}$, the measured value
is $\sigma^{J/\psi J/\psi}=5.1\pm1.0\pm1.1~\mathrm{nb}$
~\cite{LHCb}. Our calculated cross sections at LO and NLO are shown
in Fig.~\ref{scaledependence}, as functions of $\mu_r$ and $\mu_f$.
It can be seen that both $\mu_r$ dependence and $\mu_f$ dependence are
reduced at NLO level. To avoid large logarithms of
$\ln(\mu_r/\mu_f)$, as in the literature one usually estimates
theoretical uncertainties by keeping $\mu_r=\mu_f$ and varying them
from $m_T/2$ to $2m_T$. In this way, our predictions are
$\sigma_{\mathrm{LO}}=4.56\pm1.13~\mathrm{nb}$ and
$\sigma_{\mathrm{NLO}}=5.41^{+2.73}_{-1.14}~\mathrm{nb}$, which are
roughly compatible with the LHCb measured cross section.

\begin{figure}[!hbtp]
\centering
\includegraphics[scale=0.8]{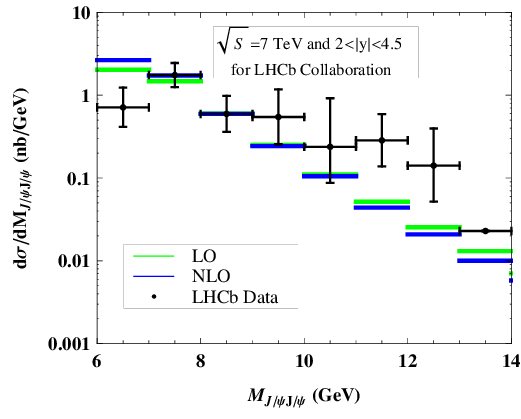}%
\caption{\small (color online). Differential cross sections in bins
of the $J/\psi$ pair invariant mass at LHCb. The data are taken from
Ref.~\cite{LHCb}. The green and blue lines denote the LO and NLO
theoretical results respectively.} \label{MassLHCb}
\end{figure}

The invariant mass distribution at LHCb is shown in
Fig.~\ref{MassLHCb}. We see that both the LO and NLO results are
inconsistent with the LHCb data, indicating that the
behaviors at both LO and NLO are very different from the LHCb data, which peaks at  small invariant mass and decreases more
slowly than the theoretical predictions at large invariant mass.
We therefore draw the conclusion that the full NLO calculation in the CS model can
not describe the LHCb data.

In the CMS conditions~\cite{CMS}:
\begin{eqnarray*}
&&|y(J/\psi)|<1.2~\mathrm{for}~p_T>6.5~\mathrm{GeV}, \mathrm{or}\nonumber\\
&&1.2<|y(J/\psi)|<1.43~\mathrm{for}~p_T>6.5\rightarrow
4.5~\mathrm{GeV},
\mathrm{or}\nonumber\\
&&1.43<|y(J/\psi)|<2.2~\mathrm{for}~p_T>4.5~\mathrm{GeV},
\label{eq:CMScut}
\end{eqnarray*}
the total cross section is measured to be
\begin{eqnarray}
\sigma_{Exp.}=1.49\pm0.07\pm0.14~\mathrm{nb}, \label{eq:CMSexp}
\end{eqnarray}
while our LO and NLO calculations for the total cross section give
\begin{eqnarray}
\sigma_{\mathrm{LO}}=0.08~\pm0.02~\mathrm{nb},~~~~
\sigma_{\mathrm{NLO}}=0.93\pm0.25~\mathrm{nb}. \label{eq:CMSresults}
\end{eqnarray}
As expected, we see the NLO calculation gives the dominant
contribution. In Eq.(\ref{eq:CMSresults}) the contribution of feeddown
process $p+p\rightarrow J/\psi+\psi(2S)+X\rightarrow 2J/\psi+X$ is
also included, which is estimated to be $30\%$ of the direct
production\cite{LO3}. Comparing Eq.(\ref{eq:CMSexp}) with
Eq.(\ref{eq:CMSresults}), we see the theoretical result is inconsistent
with the experimental data.

\begin{figure}[!hbtp]
\centering
\includegraphics[scale=0.8]{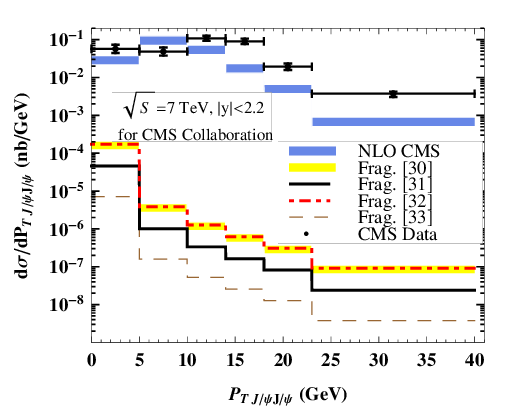}%
\caption{\small (color online). Differential cross sections in bins
of the transverse momentum of $J/\psi$ pair at CMS. The data are
taken from Ref.~\cite{CMS}. The blue band denotes the NLO results,
where the uncertainties are due to scale choices as mentioned in the
text, and the yellow band, solid, dash dotted lines represent the
fragmentation contributions including all relevant channels by three
sets of different CO matrix elements. The dashed line represents
the fragmentation contribution by taking the
${}^1S_0^{[8]}$-dominant CO matrix elements.} \label{pjjCMS}
\end{figure}

We then compare our prediction for the transverse momentum $p_{T
J/\psi J/\psi}$ distribution of $J/\psi$ pair with data. The result
is shown in Fig.~\ref{pjjCMS}. At LO, $p_{T J/\psi J/\psi}$ is
always zero, because it is a two-body final state process. At NLO,
unfortunately, as indicated in Fig.~\ref{pjjCMS}, the theoretical
result is still very different from the CMS data.
The data obviously overshoots our NLO prediction at large $p_{T
J/\psi J/\psi}$.

As mentioned before, the single parton fragmentation processes behave as
$p_T^{-4}$, which may give larger contributions at very large $p_{T
J/\psi J/\psi}$. We thus evaluate the single parton fragmentation
contribution according to Eq.~\eqref{eq:Frag}, and the results are
shown in Fig.~\ref{pjjCMS}. It can be seen that, however, the
fragmentation contribution is negligible even when $p_{T J/\psi
J/\psi}$ is as large as 30 GeV, no matter which set of CO LDMEs is
chosen. This phenomenon seems to be surprising, but actually is not
new. Similar behavior was found in Refs. \cite{chao:2011,FPE:2014} for the single $J/\psi$ inclusive production, where the
$p_T^{-6}$ contribution still dominates over the $p_T^{-4}$ contribution
even when $p_T$ is 15 times larger than the mass of $J/\psi$.
Here, the smallness of the single parton fragmentation contribution
for double $J/\psi$ production is again due to the current
experimental $p_{T J/\psi J/\psi}$ being not large enough to make fragmentation dominant. Similarly, we also do not expect the same-side-fragmentation
contribution, e.g. in Fig.~\ref{feynmandiag} (f) and (g), to be able to solve the
surplus problem for the CMS large $p_{T J/\psi J/\psi}$ data.

\begin{figure}[!hbtp]
\centering
\includegraphics[scale=0.8]{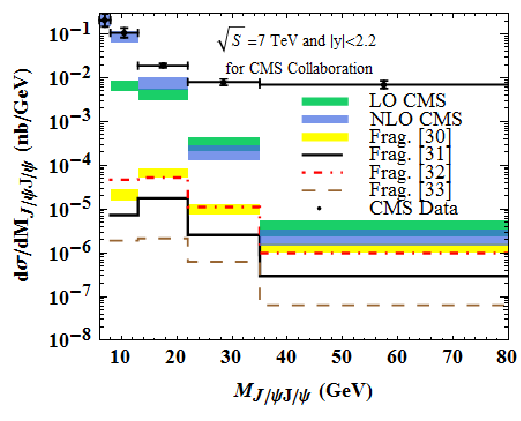}%
\caption{\small (color online). Differential cross sections in bins
of the $J/\psi$ pair invariant mass at CMS. The data are taken from
Ref.~\cite{CMS}. The green and blue bands denote the LO and NLO
theoretical results respectively, where the uncertainties are due to
scale choices as mentioned in the text. The yellow band, solid, dash
dotted lines represent the sum of the quark and gluon fragmentation
from all relevant channels by three group of different color-octet (CO) matrix
elements. The dashed line represents the fragmentation contribution
by taking the ${}^1S_0^{[8]}$-dominant CO matrix elements.}
\label{MassCMS}
\end{figure}

The invariant mass distribution at CMS is shown in
Fig.~\ref{MassCMS}. We see that the NLO result can well describe the
first two bins, but it decreases too fast beginning from the third
bin. This indicates that the behavior at NLO is very different from
the CMS data: the latter is almost flat at large invariant mass, and
larger than the NLO result by several orders of magnitude. In fact,
when $22~\mathrm{GeV}< M_{J/\psi J/\psi}< 35~\mathrm{GeV}$, the NLO
prediction is less than CMS data by almost two orders of magnitude,
and when $35 ~\mathrm{GeV}< M_{J/\psi J/\psi}<80~\mathrm{GeV}$, the
discrepancy raises to almost four orders of magnitude.

Intuitively, by examining the discrepancy in the $J/\psi$ pair mass
distribution, a large angle $J/\psi$ pair production process is
apparently needed. The quark and gluon
fragmentation processes shown in Fig.1(h) and Fig.1(i) are typically among the large angle processes. We then evaluate these fragmentation contributions, including all relevant color-singlet and color-octet channels
(Fig.1(f) and Fig.1(g) are neglected because
they are not large angle scattering processes and contribute little
to the large invariant mass distribution). The total contribution of
all concerned fragmentation channels is shown in Fig.~\ref{MassCMS}.
Unfortunately, the fragmentation contributions are found to be
negligible to the $J/\psi$ pair production, thus the discrepancy
between NLO result and CMS data can not be resolved by these
processes.

We also consider other possible sources for the discrepancy, e.g.,
the $Z^0$ boson decays to a $J/\psi$ pair: $Z^0\rightarrow
2J/\psi+X$. Under the CMS condition, the total cross section of this
process is $\sigma=2.5\times10^{-4}~\mathrm{nb}$. Its contribution
to each bin of the $J/\psi$ pair transverse momentum distribution or
invariant mass distribution is negligibly small. So the big gap
between NLO predictions and CMS data still remains.

\begin{figure}[!hbtp]
\centering
\includegraphics[scale=0.8]{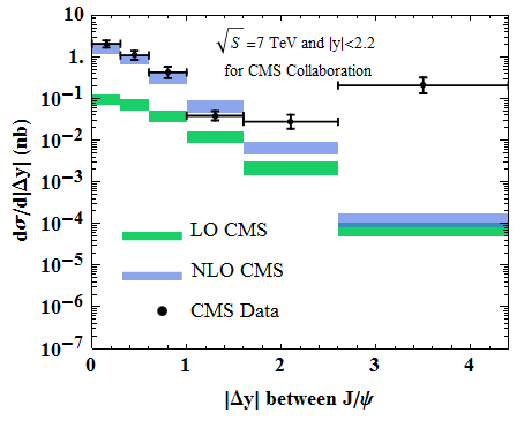}%
\caption{\small (color online). Differential cross sections in bins
of the $J/\psi$ pair $|\Delta y|$ at CMS. The data are taken from
Ref.~\cite{CMS}. The green and blue bands denote the LO and NLO
theoretical results respectively, where the uncertainties are due to
scale choices as mentioned in the text.}
\label{DyCMS}
\end{figure}

The $J/\psi$ pair rapidity difference $|\Delta y|$ distribution at
CMS is shown in Fig.~\ref{DyCMS}. We see that the NLO result can
well describe the first four bins, but it decreases too fast
beginning from the fifth bin. This is the same as the mass
distribution, because the large mass is equivalent to the large
$|\Delta y|$, and the color-singlet contributes little to a large angle
scattering process. Therefore, the fragmentation contributions are
also negligible in resolving the discrepancy between NLO result and CMS
data, so we do not label them in this figure.

\subsection{SUMMARY}

In the framework of NRQCD factorization, we
evaluate the full NLO $J/\psi$ pair production via the color-singlet channel.
We demonstrate that NLO corrections are essential for $J/\psi$ pair
production both in low $p_T$ and high $p_T$ regions, as compared to
the LO results. Our NLO calculation can give a reasonably  good
description for the total cross section observed by LHCb. However,
the NLO predictions of $p_{T J/\psi J/\psi}$ distribution, invariant
mass distribution of $J/\psi$ pair, and rapidity difference
distribution of $J/\psi$ pair are very different from the CMS data.
For the $J/\psi$ pair invariant mass distribution, the observed
flatness and orders of magnitude differences from theoretical
predictions in the large invariant mass region
($22~\mathrm{GeV}<M_{J/\psi J/\psi}<80~\mathrm{GeV}$) are hard to
explain in NLO NRQCD with color-singlet contributions, and the situation for rapidity
difference distribution is similar to the mass distribution. This strongly indicates that the CS model cannot solve the problems not only for the well known single $J/\psi$ inclusive production but also for the double $J/\psi$ production at hadron colliders.  We further take into account the contributions from quark fragmentation and gluon
fragmentation with both CS and CO channels beyond NLO in $\alpha_s$,
but find they cannot provide a sizable contribution to the large angle
production of the $J/\psi$ pair. Our calculation implies that at
low $p_T$ the color-singlet contribution may be dominant but
the color-octet contribution may be important at large
$p_T$, as shown in Ref.\cite{LOcomplete} with LO color-octet calculations. Apparently, new processes or
mechanisms are needed to simultaneously enlarge the total cross
section, improve the $p_{T J/\psi J/\psi}$ distribution, and
increase the large invariant mass distribution and large rapidity
difference distribution of $J/\psi$ pair, if the CMS data are
confirmed.

\subsection{ACKNOWLEDGMENTS}
We thank Y. Q. Ma, C. Meng, H. S. Shao, and Y. J. Zhang  for
valuable discussions and suggestions, and J.P. Lansberg for useful communications. This work was supported in
part by the National Natural Science Foundation of China (No
11475005, No 11075002), and the National Key Basic Research Program
of China (No. 2015CB856700).


\end{document}